\documentstyle[twocolumn,aps]{revtex}

\advance\textwidth 0.3in
\advance\oddsidemargin -0.15in
\advance\evensidemargin -0.15in

\input{epsf}
\begin{document}
\draft

\title{Highly Optimized Tolerance: Robustness
and Power Laws in Complex Systems}

\author{ J.M. Carlson}
\address{ Department of Physics,
University of California, Santa Barbara, CA 93106}

\author{ John Doyle}
\address{ Control and Dynamical Systems, California Institute of Technology,
Pasadena, CA 91125}

\date{\today}

\makeatletter
\def\@maketitle{%
\@preprint
\@title
\ifdim\prevdepth=-1000pt \prevdepth0pt\fi
\@authoraddress
\@date
\par
\ifdim\prevdepth=-1000pt \prevdepth0pt\fi
\leftskip=0.10753\textwidth \rightskip\leftskip
\dimen0=-\prevdepth \advance\dimen0 by17.5pt \nointerlineskip
\small\vrule width 0pt height\dimen0 \relax
\newabs\par
}

\makeatother

\long\def\newabs{%
We introduce {\it highly optimized tolerance} (HOT), a
mechanism that connects evolving structure and power laws
in interconnected systems. HOT systems arise, e.g., in biology
and engineering, where design and evolution create complex systems
sharing  common features, including
(1) high efficiency, performance, and robustness to designed-for
uncertainties,
(2) hypersensitivity to design flaws and unanticipated perturbations,
(3) nongeneric, specialized, structured configurations,
and (4) power laws.
We introduce HOT states in the context
of percolation, and contrast properties of the
high density HOT states with random configurations near
the critical point. While both cases exhibit power laws, only HOT
states display properties (1-3) associated with design and evolution.
\vskip\baselineskip

PACS numbers: 05.40.+j, 64.60.Ht, 64.60.Lx, 87.22.As, 89.20.+a
}

\maketitle
\setcounter{footnote}{0}

Evolution from primitive isolation
to more dense interconnection is
an important strategy for both biological \cite{gould}
and technological systems \cite{Pool}
as they progress towards increasing robustness and higher performance.
However, interconnections also make systems
more vulnerable to catastrophic breakdowns associated with
cascading failures initiated by seemingly innocuous local events.
Recently a great deal of attention has been given to the fact that
many complex systems share a common
statistical attribute: the distributions of sizes of events satisfy
power laws \cite{bakbook}.
Examples include the probability distributions describing
the number of fatalities and/or economic losses
due to earthquakes, hurricanes, floods,\cite{USGS,EOS}
epidemics, and social conflicts,  customers
affected by power outages \cite{power},
delays associated with traffic jams \cite{kai}, and many quantities
on the internet \cite{internet}.

In this letter we introduce a mechanism for power laws
which is relevant for systems which are
optimized, either by design or natural selection, for high
output in the presence of some external hazard.
Optimization causes systems to evolve away from generic states
towards rare, specialized configurations.
Interestingly, we find that tradeoffs between maximizing
yield and minimizing risk quite generically leads to heavy tails
(power laws) in the distribution of failure events. We refer to our
mechanism as {\it highly optimized tolerance} (HOT),
suggesting systems designed for high performance in an
uncertain environment, and operating at densities
well above the standard critical point. However, along with the high
performance comes vulnerability and brittleness with respect to design
flaws and unanticipated changes in the external conditions.

Our mechanism provides a sharp contrast with the widely
popularized alternative scenario in which open systems
evolve to a critical or bifurcation point. In that picture,
systems are said to be at the \lq\lq edge of chaos"
\cite{kauffman} or in a self-organized critical (SOC)
state \cite{bakbook}. In model systems the internal dynamics
lead a key macroscopic control parameter or density
to converge to the critical point. The system is otherwise free
to explore a wide variety of microscopic configurations which
are consistent with the specified density, and
power laws and self-similarity arise
as familiar hallmarks of criticality.

\begin{figure}[t]
\centerline{\epsfysize=3.00in \epsfbox{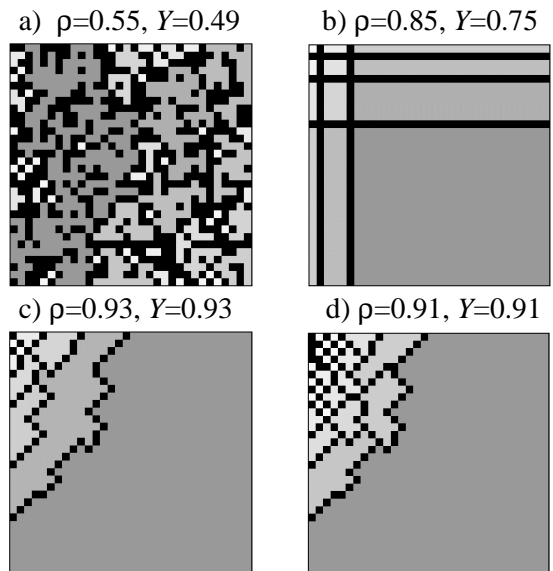}}
\caption{Sample configurations for (a) the random case near $p_c$,
(b) a HOT grid, and HOT states obtained by evolution at (c) optimal
yield, and (d) a somewhat lower density.  Unoccupied sites are black,
and clusters are grey, where darker shades indicate
larger clusters. }
\end{figure}

We focus on a very simple setting, two-dimensional site
percolation \cite{stauffer}
on an $N\times N$ square lattice. We use $N=32$ throughout
for the numerical examples, so that features of specific configurations
in Fig.~1 are easily visualized. We have verified numerically and
in some cases analytically \cite{CD1} that the power laws extend to
large $N$, where the statistics become smoother and the transitions
sharper.  In the random case
(i.e.~no design) sites are independently occupied with probability $p$
and vacant with probability $(1-p)$, so that
for a given density $\rho=p$ all configurations are equally likely.
In contrast, design implies a selection of special configurations,
in our case associated with optimization for yield in the presence of
external risk.

In the standard forest analogy, occupied sites
correspond to trees, and risk is associated with fires. The yield
${Y}$ is defined to be the average density of trees left unburnt
after a spark hits.  If a spark hits an unoccupied site,
nothing burns.  When the spark hits an occupied site the
fire spreads throughout the associated cluster, defined to be the connected
set of $c$ nearest neighbor occupied sites.   Let $f(c)$ denote
the distribution of events of size $c$, and let $F(c)$ denote the
cumulative distribution of events greater than or equal to $c$.  The
yield is then $Y(\rho)=\rho - <f>$ where the average $<f>$ is computed
with respect to both the ensembles of configurations and the
spatial distribution $P(i,j)$ of sparks.
By translation invariance, results for the random
case are independent of the distribution
of sparks, while $P(i,j)$ is a central ingredient
for the design of tolerant configurations. We assume $P(i,j)$ is
given, e.g., in terms of the past history of events.  HOT states
arise when we optimize the yield ${Y}$.

In Fig.~2a we plot yield ${Y}$ as function of the initial density
$\rho$ for a variety of different scenarios.
The maximum possible yield corresponds
to the diagonal line: ${Y}=\rho$, which is obtained if a vanishing
fraction of the sites are burned after the spark lands.
In the limit $N\to\infty$ it is possible to attain the
maximum yield for the full range of densities.
The diagonal breaks up into three regimes: a range of {\it isolated states}
composed of small, well separated clusters
at low densities, which terminates in a generic {\it critical point},
$p_c$, beyond which maximum yield is obtainable
only for a measure zero
subset of {\it tolerant} high density configurations.

The yield curve for the random case
is depicted by the dashed line in Fig.~2a, and illustrates
the isolated and critical regimes. At low densities the results
coincide with the maximum yield. Near $\rho=p_c$ there is
a crossover, and ${Y}(\rho)$
begins to decrease monotonically with $\rho$, approaching
zero at high density. The crossover becomes sharp as $N\to\infty$
and is an immediate consequence of the percolation transition,
marking the emergence of an infinite
cluster when $p=p_c$. In the thermodynamic limit only events involving
the infinite cluster result in a macroscopic event
and ${Y}(\rho)=\rho-{P_\infty}^2(p)$. Here $P_\infty(p)$ is the
percolation order parameter, i.e., the
probability a given site is in the infinite cluster.
A typical random configuration at peak yield
is illustrated in Fig.~1a. The fractal appearance of
the clusters is a key signature of criticality.

The goal of design is to push the yield towards the upper bound
for densities which exceed the critical point.
This requires selecting nongeneric (measure zero) configurations,
which we refer to as {\it tolerant states}.
We define HOT states to be those which specifically optimize
yield in the presence of a constraint (see Figs.~1b-d).
Unlike random configurations, in tolerant states the connected
clusters are regular in shape and separated by
well defined barriers consisting of closed contours of
unoccupied sites. A HOT state corresponds to a forest which is
densely planted to maximize the timber yield, with
fire breaks arranged to minimize the spread of damage.

Optimization requires that we specify $P(i,j)$ and any applicable
constraints. Precise knowledge of the position $(i,j)$ of the next spark
trivially leads to HOT states where that site is vacant, and hence no fire.
Alternately, if $P(i,j)$ is spatially uniform,
HOT states consist of regular cells of equal size (in the
thermodynamic limit). More interesting cases arises when
$P(i,j)$ is a nontrivial distribution, with regions of high
and low probability. For our numerical examples  we use:
\begin{eqnarray}
&P(i,j)= P(i)P(j)\nonumber\\
&P(x)\propto 2^{-[(m_x+(x/N))/\sigma_x]^2}
\label{Pij}
\end{eqnarray}
where $m_i=1$, $\sigma_i=0.4$, $m_j=0.5$ and $\sigma_j=0.2$.
We choose the tail of a Gaussian to dramatize that power laws
emerge through design even when the external distribution is far from a
power law. Otherwise Eq.~(\ref{Pij}) is chosen somewhat arbitrarily
to avoid artificial symmetries in the HOT configurations.

\begin{figure}[h]
\centerline{\epsfysize=1.80in \epsfbox{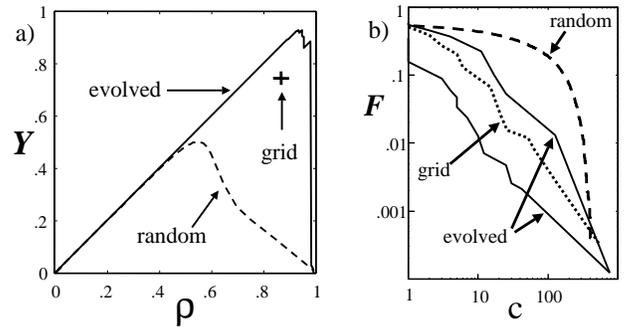}}
\caption{Comparison between HOT states and random
systems at criticality:
(a) Yield vs.~Density: ${Y}(\rho)$, and (b) cumulative
distributions of events $F(c)$ for cases (a)-(d) in Fig.~1. }
\end{figure}

Specific constraints make the optimization procedure tractable.
The HOT configuration illustrated in Fig.~1b is obtained
subject to a constraint in which the lattice is fully occupied except
for horizontal and vertical lines of vacant sites or \lq\lq cuts"
which divide the system into rectangular clusters.
In this case it is straightforward to determine the constrained, global
optimum by searching over the number and locations of cuts.
Analytical calculations can be made in the thermodynamic limit \cite{CD1}.
Solutions for Cauchy, exponential, and Gaussian
distributions $P(i,j)$ in addition to Eq.~(\ref{Pij}), exhibit power
law tails in the distribution of burn events, where for
a broad class of distributions the scaling is asymptotically
independent of  $P(i,j)$. Numerical results for
the case of Eq.~(\ref{Pij}) are illustrated in Fig.~2b.
The key point is that in the tolerant regime power laws events
are highly generic for a variety of (not necessarily power law)
input distributions. In HOT states resources (in this case the
cuts) are concentrated around regions
of high $P(i,j)$, creating small clusters, while few resources
are spent where  $P(i,j)$ is small, creating large clusters.  Since the
probability of events $f(c)$ is the sum of all the $P(i,j)$ in  clusters of
size $c$, optimizing yield balances cluster size and
probability, which produces power law tails.

There are many
alternative optimization schemes  associated with different
constraints.  Next we turn to a local and incremental
algorithm,
which is reminiscent of evolution by natural selection.
We begin with an empty lattice, and add grains one at a time to
sites which maximize expected yield at each step.
For asymmetric $P(i,j)$ such as Eq.~(\ref{Pij})
this algorithm is deterministic.  We obtain a sequence of
configurations of monotonically increasing density, which passes through
the critical density $p_c$ unobstructed.  Indeed, $p_c$ plays no special
role. At much higher densities there is a maximum yield point
followed by a drop in the yield. The yield curve  ${Y} (\rho)$ is plotted
in Fig.~2a for the $P(i,j)$ given in Eq.~(\ref{Pij}).

A sample HOT configuration generated by this algorithm is illustrated
in Fig.~1c for a density near the maximum yield point in
Fig.~2a. This optimization  explores only a small fraction of the
configurations at each density $\rho$.  Specifically,
$(1-\rho)N^2$ of the $N^2 \choose {(1-\rho)N^2}$
possible configurations are searched.  Nonetheless,
yields above $0.9$ are obtained on a $32\times 32$
lattice, and in the thermodynamic limit the peak yield approaches
the maximum value of unity. While the clusters are not
perfectly regular, the configuration has a clear cellular pattern,
consisting of compact regions enclosed by well defined barriers.
As shown in Fig.~2b, the distribution of events $F(c)$ exhibits a power law
tail when $P(i,j)$ is given by Eq.~(\ref{Pij}).  This is the case for
a broad class of $P(i,j)$, including Gaussian, exponential, and Cauchy.

Interestingly, in the tolerant regime our algorithm produces power
law tails for a range of densities below the maximum yield, and
without ever passing through a state that resembles the (fractal) critical
state. This is illustrated in Figs.~1d and 2b where we plot the event size
distribution $F(c)$ (lower of the \lq\lq evolved" curves) for a density
which lies below that associated with the peak yield.   Note that this
configuration has many clusters of unit size $c=1$
in checkerboard patterns in the region of high
$P(i,j)$ in the upper left corner. The fact that power
laws are not a special feature associated with a single density is in
sharp contrast to a traditional critical phenomena.

Like criticality, HOT states display certain \lq\lq universal" features, and
the scaling properties are determined by limited sets of  key variables.
These variables are different in the two cases, but
the most interesting properties of HOT states
are those which even more clearly distinguish tolerance from
criticality. In contrast to the fractal percolation clusters,
regions in the HOT state are both regular and structured depending
on the $P(i,j)$, and also highly sensitive to changes in
$P(i,j)$. For example, if a configuration which is optimized
for a Gaussian distribution is
subjected to a uniform distribution of hits, then the
distribution of events {\it increases} with size: $f(c)\sim c$ \cite{CD1}.
For the random critical case the event size distribution is {\it a priori}
independent of the spark distribution.

For a given density the expected event sizes associated
with HOT states are much smaller than those of random configurations.
In Fig.~2b the random case exhibits
the flattest distribution with events of the largest average size,
in spite of the fact that it corresponds to the lowest density.
However, there is a robustness tradeoff which
introduces new sensitivities in the HOT state which are not
present in random cases. For example, the HOT state is extremely
sensitive to design flaws. If an element in the surrounding
barrier of vacant sites is absent (that is, occupied by a tree),
then fire leaks through the barrier into the surrounding regions.
In contrast, in random configurations small changes 
do not alter the distribution of events. Robustness to additional
uncertainties such as design flaws or multiple sparks can be designed for
at some cost in yield. A common engineering design strategy is to
simply back off from the peak yield (e.g.~Fig.~1c), and consider a
configuration more analogous to that illustrated in Fig.~1d.

In summary, the distinguishing features of the HOT state include
(1) high yields robust to designed-for uncertainty,
(2) hypersensitivity to design flaws and unanticipated perturbations,
(3) stylized and structured configurations, and
(4) power law distributions.
Percolation seems to be the simplest  template for introducing
HOT states and contrasting their properties with criticality.  For a more
unified perspective, the
random case can be viewed as a very primitive design with density as
the {\it only} design parameter.  In this case, the critical point coincides
with the maximum yield, making this a natural alternative to SOC whereby
primitive systems might evolve to criticality.  More importantly, adding
even modest levels of additional design
moves yields well above the random critical point.
While both HOT states and critical points  exhibit power laws, this is
the least consequential of the four noted HOT features.

This simple model is emphatically not meant
to realistically represent any specific system, and is at best remotely
connected  with  forest management. Nevertheless, it is striking
how commonly complex systems have {\it all} the features of
the HOT state. We  briefly review a few of these systems, particularly
those previously studied emphasizing power laws and criticality, in order
to underscore that  power law statistics alone should not necessarily be
interpreted as signatures of criticality.  For example, in this context,
highway traffic is widely studied \cite{kai}. However,
the complete highway system is highly structured with throughput
dominated by design, including multiple and specialized lanes and ramps,
the use of buses and vans, and perhaps most importantly, drivers capable of
sophisticated active feedback control and collision avoidance. Traffic
flow can also be hypersensitive to, say, accidents that block lanes.

Modern computer networks, which exhibit many power laws \cite{internet},
are also highly
structured with routers, caches, and sophisticated multilayer protocols to
provide high throughput. The network is robust to moderate variations in
traffic, or loss of a router or line, but extremely sensitive to bugs
in network software, underscoring the importance of software reliability.
The Ariane 5 crash and Y2K problems are a few
other examples of our vulnerability in software intensive systems.
The high connectivity and throughput of the electric power network, plus
protective relay control, provide great robustness to most perturbations,
but can also lead to large cascading multibillion-dollar failures from
apparently small and innocuous initiating events \cite{power}.

Biological systems
show extreme robustness at all levels, including cells, organisms, and
ecosystems, but also have hypersensitivity to the alteration of a single
gene or specie. Despite possibly fractal signaling and transport structures,
which are themselves products of design,
the overall hierarchy of organism, organs, cells, organelles, and
macromolecules is highly structured. Finally, design plays a surprisingly
large role even in
natural disasters such as hurricanes, earthquakes, floods, and tornadoes.
According to a recent report \cite{EOS}, economic
losses associated with natural disasters are on the rise due to
an increased concentration of population and infrastructure in
high risk areas. Thus the authors argue that with respect to economic
losses,
``most natural disasters are not random acts, but rather the direct
and predictable consequence of inappropriate land use.''

While these examples share the four basic features of the HOT state,
it can be difficult to precisely characterize and quantify
the role of design in complex technical and biological systems without going
into great detail. Thus in any system there may be confusion as to which
feature are due to design and which are due to dynamical or statistical
mechanisms more familiar in physics.
In advanced systems, designed features are so dominant and pervasive that we
often take them for granted.   While generic complexity emerges from a
featureless substrate,  the complexity in designed systems often leads to
apparently simple, predictable, robust behavior.  As a result, designed
complexity becomes increasingly hidden, so that its  role in determining the
sensitivities of the system tends to be underestimated by nonexperts,
even those scientifically trained. Furthermore, because HOT systems are
simultaneously robust and sensitive to their components and environment,
it is difficult to predict a priori which details are important.

Nonetheless, the special sensitivity of  HOT systems
can lead to methods for
separating features associated with design from those
which collectively and self-consistently emerge from internal
processes. When the behavior of a system changes radically
in response to small rearrangements, variations
in the boundary conditions, or  the replacement of
highly nongeneric (measure zero) elements  with
more random versions, then, chances are, the system is HOT.
Evidence for HOT states can even be found in familiar laboratory
experiments.  For example, water flowing in structured (straight,
smooth) pipes can be laminar to Reynolds numbers of $10^5$. For
the same pressure drop, this results in flows which are
much greater than for more generic (e.g. rough, twisting, turning) pipes,
which become turbulent at Reynolds numbers below $10^3$. However,
``designed'' pipes are hypersensitive to microscopic details, such as
small concentrations of polymers,  wall roughness, or vibrations.

Random analogs of familiar systems in engineering and biology
are so obviously different from our daily experience that the comparison
is almost absurd.  A truly random traffic system
would have no lanes, dividers, traffic laws, collision avoidance or other
control systems, and would exhibit a \lq\lq phase transition" of sorts
at very low densities compared to  the standard operating conditions
of our current highways. It has been the work of
auto makers, civil engineers, and city planners   to
continue developing new methods to increase throughput without
exceeding a maximum yield point by incorporating increasingly
structured and sophisticated features into each subsystem, if not
globally optimizing the design. Computer networks, power grids,
and biological systems are even more highly structured, with hierarchies,
protocols, and enormous amounts of feedback.
The message we extract from our simple percolation model
is that in these systems a detailed representation of
the internal interactions  studied in a generic setting
may be much less accurate than a  coarser representation  coupled with
better characterizations of nongeneric elements
such as perturbations and boundary conditions.
In the HOT state design influences the most basic
properties, and therefore must be  taken into account
throughout modeling, analysis, and simulation.

\acknowledgements
This work was supported by the David and Lucile
Packard Foundation, NSF Grants No.~DMR-9212396 and DMR-9813752,
and a DOD MURI grant for
\lq\lq Mathematical Infrastructure for Robust Virtual Engineering."

\end{document}